\documentclass[12pt]{article}
\usepackage{t1enc}
\usepackage[latin1]{inputenc}
\usepackage [dvips]{graphicx}
\usepackage{flafter}
\usepackage[dvips]{rotating}
\usepackage{latexsym}
\begin{document}
\begin{center}{\bfseries \large  Does telomere elongation in cloned organisms lead to a longer
 lifespan if cancer is considered?}
\end{center}
\centerline{Michael Masa$^1$, Stanis{\l}aw Cebrat$^2$ and Dietrich Stauffer$^1$}
\bigskip
\noindent
$^1$ Institute for Theoretical Physics, Cologne University, D-50923 K\"oln,
Euroland \
\noindent
$^2$ Institute of Genetics and Microbiology,
University of Wroc{\l}aw,\\ ul. Przybyszewskiego 63/77, PL-54148 Wroc{\l}aw,
Poland \
\\ \\ 
\begin{abstract}
As cell proliferation is limited due to the loss of telomere repeats in DNA of normal
somatic cells during division, telomere attrition can possibly play an important role in
determining the maximum life span of an organism. 
With computer simulations of cell culture development in organisms, which consist of
tissues of normal somatic cells with finite growth, we otain an increase of life span for
longer telomeric DNA in the zygote.
By additionally considering a two-mutation model for carcinogenesis and indefinite proliferation by
the activation of telomerase, we demonstrate that the risk of dying due to cancer can outweigh 
the positive effect of longer telomeres on the longevity.
\end{abstract}
\begin{normalsize}
\end{normalsize}
\section{Introduction} 
Telomeres are tandem repeated noncoding sequences of nucleotides at both ends of the DNA in
eukaryotic chromosomes stabilizing the chromosome ends and preventing them
from end-to-end fusion or degradation \cite{blackburn}. Polymerase cannot completely replicate
the 3' end of linear DNA, so telomeres are shortened at each DNA replication \cite{harley}.
This end replication problem leads to a finite replicative capacity for normal somatic cells
\cite{olovnikov}.
They can only divide up to a certain threshold, the Hayflick limit \cite{hayflick01}.
The enzyme telomerase, repressed in most normal somatic cells, synthesizes and elongates
telomere repeat sequences at the end of DNA strands so that certain cells like germline
cells are immortal and indefinite in growth \cite{greider,shay01}.

Most forms of cancer follow from the accumulation of somatic mutations \cite{nordling}.
Cancer-derived cell lines and 85-90\% of primary human cancers are able
to synthesize high levels of telomerase and thus are able to prevent further shortening of
their telomeres and proliferate indefinitely \cite{kim}. 
But if cells are premalignant or already cancerous and telomerase is not yet activated,
the proliferation of these cells and the accumulation of mutations is determined by
their further replicative capacity \cite{moolgavkar01}.
So the frequency of malignant cancer should be higher for longer telomeres.

Recently published data show that longer telomeric DNA increased the life span of nematode worms
\cite{joeng}. So there may be a positive effect on the longevity of complex cloned organisms
with renewing tissues if the telomere length in zygote cells is increased \cite{lanza}. 
As the probability for the incidence of cancer is correlated with the replicative potential of the
mutated cells \cite{shay02}, one can ask the following question: 
Is an extension of lifespan of cloned organisms possible if telomeres in embryonic cells are
elongated and cancer is considered?
An answer to this question could be given by computer simulations as the presented model focuses
on ageing by the loss of telomeres in DNA.

After presenting the basic model of telomere attrition in organisms and computation results,
we explain how cancer and telomerase are introduced and discuss the effects of different initial
telomere lengths.

\section{Basic model of biological ageing due to telomere shortening}
\begin{figure}[!ht]
\centering
\includegraphics{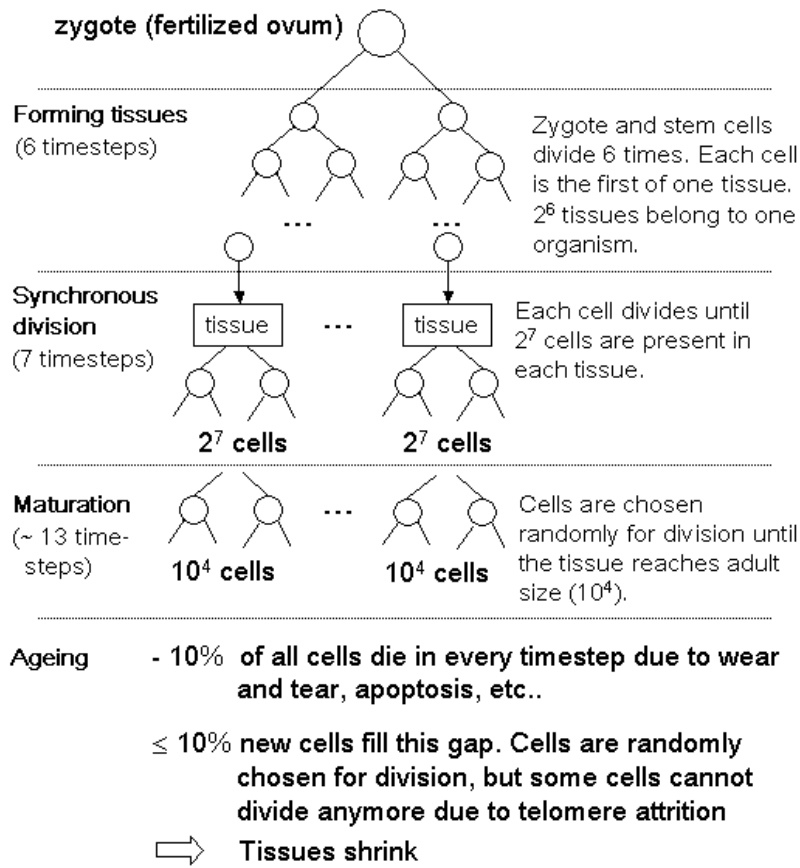}
\caption{Dynamics of cell proliferation in one organism in the basic model}
\label{bild01}
\end{figure}

As shortening of telomeres is one of the supposed mechanisms of ageing on cellular level,
there are many different approaches to model telomere loss \cite{levy01,aviv,olofsson,sozou}.
In our basic model every organism is developed from a single progenitor cell, the zygote
(figure \ref{bild01}).
The initial telomere length of zygote cells is assumed to be normally distributed with mean
$\mu_{z}$ and standard deviation $\sigma_{z}$ \cite{buijs}.
Telomere repeats lost per division (TRLPD) are randomly chosen at each division of every cell
from a normal distribution with mean $\mu_{TRLPD}$ and standard deviation $\sigma_{TRLPD}$.
A dividing cell produces a clone who inherits the replicative capacity of the progenitor
cell at this age.
All normally distributed variables are generated with the Box-Muller method \cite{chinellato}.
Cells can divide until nearly all their original telomeres are lost.

For every organism the dynamics of the model is as follows: Divisions of the zygote
and the stem cells derived from it occur 6 times in the early embryo.
Each of these cells is the progenitor of one tissue. This is followed by a period 
of population doublings where all cells divide once in every timestep
until $2^7$ cells are present in each tissue.
In the following maturation stage, cells are chosen randomly for division until each tissue
reaches the adult size of $10^4$ cells. It takes about 26 timesteps until an organism is mature.

Ageing starts now, cells first die with 10\% probability due to events like necrosis or apoptosis.
10\% of the cells of the corresponding tissue are then randomly chosen for division to fill
this gap. The replacement does not have to be complete as the chosen cell could probably not divide
anymore due to telomere attrition. After some time the tissue will start shrinking. 
The random choice of dying and dividing cells in differentiated tissues is in accordance with
nature as for example in epithelium the choice of cells to be exported from the basal layer is
random \cite{cairns,frank}. 
The organism dies, if its actual size reaches 50\% of its mature size.
\subsection{Results without cancer}
\begin{figure}[!ht]
\centering
\includegraphics[width=8cm,angle=-90]{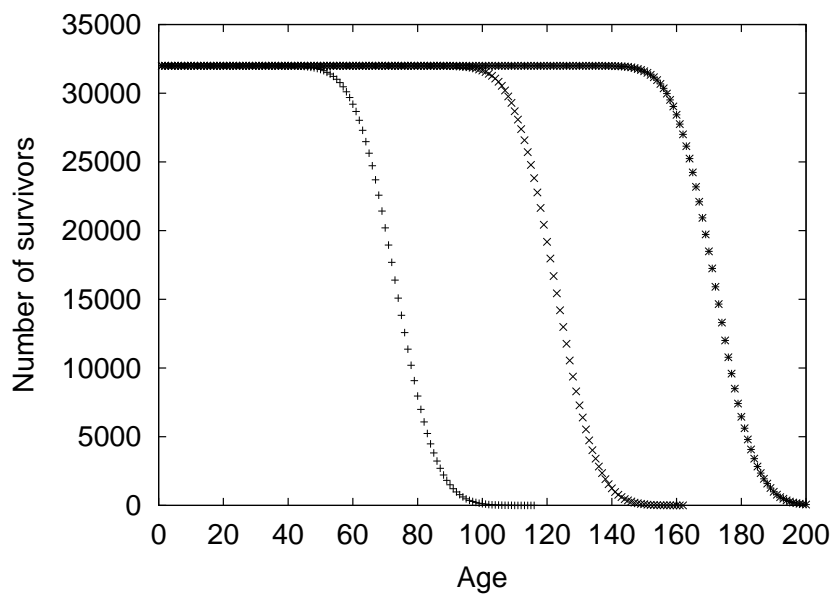}
\caption{Age distribution of 32000 organisms with telomere lengths of $\mu_z=1500(+)$,
 $\mu_z=2000$($\times$) and $\mu_z=2500(*)$; $\sigma_z=100$, $\mu_{TRLPD}=50$, $\sigma_{TRLPD}=10$}
\label{bild02}
\end{figure}

Age distributions for different mean telomere lengths in the zygote cells are shown
(figure \ref{bild02}).
The shape of these distributions is very analogous to empirical data of many human and animal
populations. We also obtain a positive effect on the longevity of the organisms if the mean
telomere length in the precursor cell is increased.

The chosen mean doubling potentials for primary cells are 30, 40 and 50 with the choice
of $\mu_z=1500,2000,2500$ and $\mu_{TRLPD}=50$. The number of mitotic divisions observed
in human fibroblasts is higher \cite{allsopp01}, but the choice of this parameters
is reasonable as the number of considered cells per mature organism (640000) in this
model is also much lower than in human organisms where the total number of cells is of the order
of $10^{13}$ \cite{cairns}.
\section{Introducing carcinogenesis and telomerase}
Clonal cancer is now introduced in the model. In accordance to the model of Moolgavkar et al.,
one of our assumptions is that malignant tumors arise from independently mutated progenitor cells
\cite{moolgavkar02}.
For most forms of carcinoma, transformation of a susceptible stem cell into a cancer cell is
suggested to be a multistage process of successive mutations with a relatively low probability
for the sequential stages \cite{nordling,armitage01}.
Two independent and irreversible hereditary mutation stages are considered here, which can occur
at every level of development of the organism during cell division.

The first premalignant stage is a promotion: A dividing cell can mutate with small
probability $p_{mut}$ \cite{cairns}. All descendant cells inherit this mutation.
This mutation leads to a partial escape from homeostatic control of growth by the local cellular
environment \cite{moolgavkar01,sarasin}.
Cells on the promotion stage have a selective advantage over unaffected cells \cite{armitage02}.
In our model they are chosen first for division during maturation and for filling up the gap in
the ageing period.

The subsequent transition can occur again with probability $p_{mut}$ during division. If a cell
reaches this second stage of mutation it is a progenitor of a carcinoma.
An explosive clonal expansion to a fully malignant compartment happens \cite{sarasin}.
This cell and the clonal progeny doubles in the current timestep until it is no more possible
due to telomere attrition.
This expansion leads only to an increase of the malignant cell population size.
As a certain fraction of cells is killed per unit time and clonal expansions only occur with a
very small probability, the tumor enviroments may not continue growing, eventually shrink,
or even die \cite{hiyama02}. We assume that it is necessary for advanced cancer progression and
therefore for the development of a deadly tumor that fully mutated cells are able to
activate telomerase \cite{shay03}. 

In our model, telomerase activation is possible at every age of the organism in normal and
mutated cells during division with a very low probability $p_{telo}$. The irreversible loss
of replicative potential is stopped in these cells.
As the contribution of telomerase to tumorigenicity is not yet completely understood,
\cite{hiyama,kanjuh,blasco}, we assume that death of an organism due to cancer occurs if
telomerase is reactived in at least one fully mutated cell \cite{moolgavkar03}. 
We treat the time interval between the occurence of a deadly tumor and death due to it as
constant, so we set this interval to zero.
In our carcinogenic process there are two ways to reach the critical stage of indefinite
proliferation of immortal cancerous cells:
\begin{enumerate}
\item  Telomerase is already reactivated in a cell which reaches the second mutation stage.
\item  Activation of telomerase occurs in one fully mutated cell during the explosive clonal
       expansion of cancerous cells after a cell has reached the second stage of mutation.
\end{enumerate}
\subsection{Effects of different telomere lengths considering cancer}
\begin{figure}[!h]
\centering
\includegraphics[width=8cm,angle=-90]{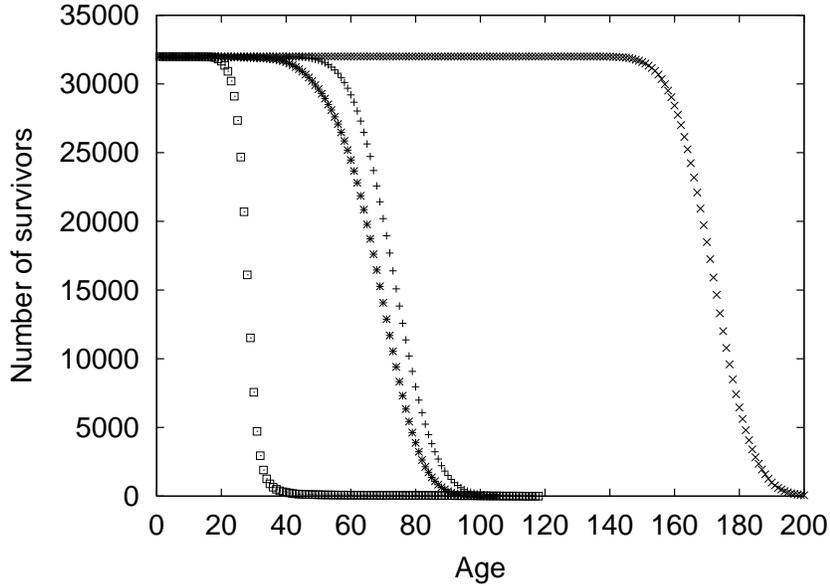}
\caption{Age distribution of 32000 organisms with telomere lengths of $\mu_z=1500$ with ($*$) and without cancer (+) and $\mu_z=2500$ with ($ \Box $) and without cancer ($\times$); $\sigma_z=100$,
$\mu_{TRLPD}=50$, $\sigma_{TRLPD}=10$. Cancer mutations are possible with $p_{mut}=5*10^{-5}$.
 Telomerase can be activated with $p_{telo}=10^{-5}$.}
\label{bild03}
\end{figure}

Figure \ref{bild03} shows simulation results for $\mu_z=1500$ and $\mu_z=2500$. 
 As we considered a lower complexity by chosing a lower number of tissues and cells per
organism, we assumed higher mutation rates for the incidence of cancer than observed in 
nature \cite{luebeck,drake}.

The age distribution for shorter initial telomere lengths considering cancer is shifted to
the left but still very old organisms exist. Cancer increases only the probability to die in a
certain range of age which varies for different telomere lengths. For longer telomeres the age
distribution is again shifted to the left but even behind the distributions for shorter telomeres 
with and without considering cancer.
There is a strong increase in the risk of dying because of cancer and hence the life
expectation is much lower than for shorter initial telomeres.

\begin{figure}[!ht]
\centering
\includegraphics[width=8cm,angle=-90]{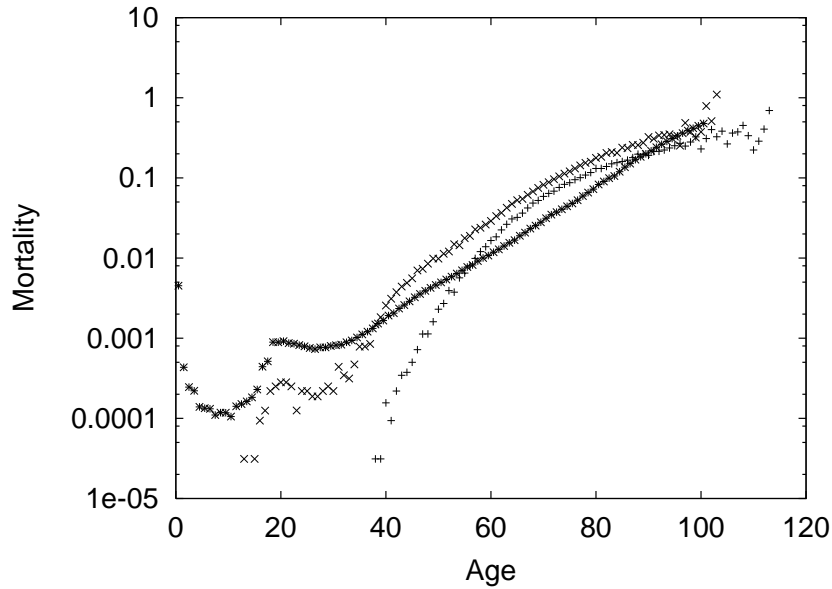}
\caption{Mortality function for $\mu_z=1500$ with ($\times$) and without cancer (+), 32000 organisms considered, $\sigma_z=100$, $\mu_{TRLPD}=50$, $\sigma_{TRLPD}=10$; German men ($*$) from www.destatis.de (June 2004) (Sterbetafel 2000/2002)}
\label{bild04}
\end{figure}

The force of mortality resulting from this model is shown for $\mu_z=1500$
with and without considering cancer in comparison to empirical human mortality 
data (figure \ref{bild04}). 
It agrees very well with human mortality functions 
provided cancer is incorporated into the model, 
and shows a mortality deceleration at advanced age, 

\section{Conclusion}
The expected simulation result of the basic model without cancer is an increase of life span of
most organisms with longer initial telomeres. After introducing somatic
mutations promoting cancer and telomerase activation in this model, the survival probability
is lower for each considered initial telomere length in certain time intervals in adult ages.

But even low probabilities for the two mutation stages and for the activation of telomerase lead
to a strong reduction of life span for longer telomeres. 
So the implication of two-stage carcinogenesis for the incidence of cancer in this simple model
of cell proliferation in organisms is that the life span of complex cloned organisms cannot be
increased by artificially elongating telomeres in primary cells.

Further improvements, extentions and applications of this model are possible. With respect to
the role of telomeres and telomerase in carcinogenesis, maybe this computational approach can
contribute to the development of a comprehensive theoretical model in oncology uniting mutagenesis
and cell proliferation \cite{gatenby}.
\\ \\
{\bf Acknowledgements}
\\
We wish to thank the European project COST-P10 for supporting visits of MM and DS 
to the Cebrat group at Wroc{\l}aw University and the Julich supercomputer center
for computing time on their CrayT3E. CS was supported by Foundation for Polish
Science.

\end{document}